\documentstyle[11pt,aaspp4,tighten]{article}

\lefthead{Graham {\it et al.}}
\righthead{VIRIS}

\begin{document}

\title{VIRIS: A VISUAL-INFRARED IMAGING SYSTEM FOR THE LICK OBSERVATORY
       1-M TELESCOPE}

\author{James R. Graham and Lynne A. Hillenbrand$^a$}
\affil{
       Department of Astronomy\\
       601 Campbell Hall\\
       University of California\\
       Berkeley, CA 94720\\
       (jrg@astro.berkeley.edu, lah@astro.caltech.edu)\\
       }

\author{Anthony A. Misch}
\affil{
       UCO/Lick Observatory\\
       University of California\\
       Santa Cruz, CA 95064\\
       (tony@ucolick.org)\\
       }


\slugcomment{Accepted: Publications of the Astronomical Society of the Pacific}

\vfil

\begin{abstract}
We describe a system in use at the Lick Observatory 1-m Nickel
telescope for near-simultaneous imaging at optical and near-infrared
wavelengths.  The combined availability of a CCD and a NICMOS-3 camera
makes the system well-suited for photometric monitoring from
0.5-2.2$\micron$ of a variety of astrophysical objects.  Our science
program thus far has concentrated on studying variability trends in
young stellar objects.
\end{abstract}


\vfil

\noindent 
$^a$ Current Address: Division of Physics, Math, \&
Astronomy, California Institute of Technology, Pasadena, CA 91125

\section{Introduction}

There is broad need in astronomy for simultaneous optical-infrared
photometric capabilities which enable temporal monitoring.  Examples
include recording supernova light curves, monitoring variability of
active galactic nuclei, and measuring the spectral energy distribution
of young stars. Because supernovae are standard candles, their light
curves are important and have maximum utility if well sampled and
cover multiple wavelengths.  Dense temporal sampling is valuable
because this helps to establish the epoch of maximum light, a task
that is simplified for type Ia events if $J$-band data are available,
and some immunity to the effects of dust can be achieved if
observations extend into the $K$-band where extinction is an order of
magnitude lower than at visible wavelengths (Elias \& Frogel 1981).
Variability of AGN such as quasars and Seyfert nuclei provides an
opportunity to investigate the nature of the central engine, and the
distribution and properties of dust grains via reverberation mapping
(e.g., Nelson 1996). Part of the data for Nelson's (1996) study came
from a dual optical-IR imager on the UCLA 24-inch Telescope (Nelson et
al. 1997). Another example of a simultaneous optical-IR system is the
$IJK$ camera used for the DENIS survey (Copet et al. 1997).

Our primary interest thus far has been to pursue an optical-infrared
photometric monitoring program to investigate and characterize
variability trends in the spectral energy distributions (SEDs) of
young stars.  Circumstellar accretion disk models matched to the
observed SEDs of young pre-main sequence stars are used to attempt
understanding of the complex accretion processes in these systems.  A
major uncertainty in the SED modeling procedure, however, is the
continuum variability which occurs at all wavelengths out to at least
3.5 $\mu$m.  Changes in flux of several tens to several hundreds of
percent occur on time-scales of a few hours to a few days, depending on
the source and on the wavelength regime.  There are several possible
causes for the observed monochromatic and color variability: 1)
variable stellar flux; 2) variable extinction; and 3) variable
properties of the circumstellar material.  Sufficient optical and
near-infrared monitoring has been carried out for a few sources so
that we are able to investigate their complicated behavior in
color-color and color-magnitude diagrams ({\it c.f.} Skrutskie
et al. 1996; Herbst, Herbst, \& Grossman 1994).  These patterns
seem inconsistent with simple variability explanations like changes in
intrinsic stellar flux or line-of-sight obscuration (although these
causes can not yet be discounted entirely).  Thus it may be that the
observed variability trends are due to changes in the physical
properties of the hot circumstellar gas and dust, e.g., temperature,
opacity, or geometry variations taking place in accretion columns, in
the inner disk, or in the stellar/disk wind.  Testing this hypothesis
requires a more extensive photometric dataset than currently exists.
Relevant monitoring time-scales might encompass: the dynamical time for
a wind (R$_*$ / v$_{wind} \sim$ hours), the free-fall time from the
inner disk to the stellar surface ($\sqrt{r^3/GM}\sim$ tens of hours),
the dynamical time for the disk ($1/\Omega = \sqrt{r^3/GM}\sim$ tens
of hours to years for r = 0.1-1AU), the dust destruction / formation
time ($\sim$ years), and the viscous accretion time-scale
($\alpha^{-1}~r^2/h^2~\tau_{rot}\gtrsim$ tens of years).

Both scientific and practical considerations shaped our plans for
implementing simultaneous optical-infrared measurements at the
University of California's Lick Observatory on Mt Hamilton. Our need
to image complex fields with multiple and sometimes extended sources
ruled out a single-channel photometer. However, as cost precluded
development of new cameras, combining existing instruments to meet our
broadband imaging requirements presented an attractive design
strategy.  This approach also permitted us to take advantage of
existing electronic and software controls, and to adopt existing
mechanical telescope interfaces with only minor modifications.

Since the science goals require intensive photometric monitoring of
relatively bright targets, we concentrated our instrument development
effort on the 1-m Nickel telescope, rather than the heavily
over-subscribed 3-m Shane telescope (though, in principle, VIRIS could
be used with either). We thus arrived at the option of combining
Lick's 1-m facility CCD camera and the facility IR camera, used at
both the 1- and 3-m telescopes.

\section{Optical \& Mechanical Configuration}

VIRIS is an opto-mechanical system for combining the Lick Observatory
facility NICMOS-3 camera, LIRC-2 (Gilmore, Rank, \& Temi 1994) with a
standard CCD dewar containing a 2048x2048 detector with 15 $\mu$m
pixels manufactured by Orbit, and an interface box which includes
automated filter and aperture wheels.  The use of an optical/IR
dichroic beam-splitter and a transfer lens delivers parfocal optical
and IR images to the respective dewars, and hence provides the
opportunity to perform simultaneous imaging in the optical and
infrared.
 
The LIRC-2 telescope interface mounts the side-looking infrared camera
dewar on the 1-m or 3-m telescopes at Lick Observatory.  The beam from
the secondary is relayed to the IR camera by a 45$^\circ$ IR
reflecting gold dichroic beam-splitter.  In the original version of
the LIRC-2 interface, optical light ($\lambda < 1 ~\mu$m) passes
through the dichroic to a bore-sight CCD acquisition and guide
camera. This configuration suggested that the acquisition and guide
camera could be replaced with a science grade CCD. The existence of an
alternate acquisition/guide camera (the TUB) located upstream of the
dichroic in the light path means these capabilities are not lost by
the replacement with the CCD.

LIRC-2 reimages the telescope focal plane from a position about 50 mm
upstream of the dewar window onto a NICMOS-3 detector, hence, the
focal plane lies only 200 mm behind the dichroic, and it is not
possible to locate the CCD interface box close enough to this mirror
to get simultaneous in-focus images on IR and optical detectors. We
decided that the most economical way to achieve this was to locate a
diverging lens in the beam, just below below the dichroic. This
solution throws the focal plane back, while adjustment of the lens
position provides a mechanism so that the IR camera and the CCD can be
in focus simultaneously.  The lens we decided to use is a stock
plano-concave anti-reflection coated borosilicate glass (BK7) singlet
with a -125 mm focal length from Edmunds Scientific Catalog (Part
Number M 45039). This transfer lens throws the optical focus back by a
nominal 140 mm.  

In normal operation with LIRC-2 the back focal length of the telescope
is increased by displacing the secondary from its nominal position for
a classical Cassegrain configuration by 35 mm towards the
primary. This introduces spherical aberration which is partially
compensated for at the optical focus by orienting the transfer lens
with its concave surface towards the secondary.  The dominant residual
aberration of the telescope and transfer lens combination is lateral
color amounting to $0.''3$ for the $R$ band filter at a field angle of
$1'$.  Typical seeing at the 1-meter telescope is about $1.''5$ FWHM,
and therefore this aberration is negligible.  The disadvantage of this
configuration is that the effective focal length at the CCD is
increased by a factor of 2.77, and a 15 $\mu$m pixel projects to
$0.''064$ on the sky. With this small pixel scale the readout is
rebinned on-chip 4x4 to yield a 512x512 array of $0.''26$ pixels with
a square field of view of $2.'2\times 2.'2$ which matches well the IR
field.  Focus differences between V, R, and I filters amount to no
more than $0.''2$ of image blur, and refocusing between wavelengths is
unnecessary.

The transfer lens is mounted in a finely threaded lens barrel and is
coupled to a micrometer head via a toothed timing belt, permitting a
total travel of 25 mm; the corresponding motion of the focal plane is
170 mm and hence this assembly permits easy access to adjust the lens
as well as a reproducible way of recording the focus setting.
Focusing the lens makes the optical and IR cameras
parfocal. Subsequent focus variations, primarily due to thermal
expansion and contraction of the telescope truss henceforth are handled by
changing the secondary mirror focus without further need to adjust the
transfer lens.  Experience observing with VIRIS has shown that the
transfer lens focus position is determined once at the beginning of a
run and thereafter requires no further adjustment.

The LIRC-2 dichroic has a thick gold coat; it was designed to minimize
infrared reflection losses, while having sufficient transmission to
permit bore-sight guiding at visible wavelengths.  The transmission
losses due to the dichroic are 2.5 mag at $V$, 3.5 mag at $R$, and 4.0
mag at $I$.  Since many of our targets are highly extincted and faint
at optical wavelengths, the low efficiency of the dichroic added
unacceptably large observing overheads.  One option would be to
replace the dichroic with one that has a thinner gold coat. This
choice was unacceptable since substantially improved transmission at
$I$-band could only be achieved by sacrificing reflectivity at
$J$. High performance, multi-layer, dielectric beam splitters are
available, but they are expensive.  We decided that removing the
dichroic from the optical path, by locating it on remotely
controllable translation stage, would provide the most satisfactory
solution.  This scheme has the added advantage to regular users of
LIRC-2 of improving the sensitivity of the bore-sight camera to aid
the identification of optically faint targets.  Thus, although truly
simultaneous optical and infrared observations are feasible (but see
\S \ref{eas}) our data have been obtained sequentially, with movement
of the dichroic out of and in to the beam for the optical and infrared
observations, respectively.  Furthermore, the need for short exposures
and dithering in the IR make genuine parallel optical-IR impractical.
We did not attempt to make the bore-sights of the optical and IR
cameras coincide exactly since any offset can be accommodated by a
small telescope motion.

\section{Electronics \& Software}
\label{eas}

For operation with VIRIS, the CCD camera and LIRC-2 retain their
original control electronics and Lick data acquisition systems. The
CCD uses a Lick in-house array controller with its interface running
on an Integrated Solutions 68030-based workstation. LIRC-2 uses a
Leach controller run from a Sun Microsystem Sparc-5. Both machines run
UNIX and X-11 and are fully networked.

Both data acquisition systems obtain telescope status information for
their respective image headers by querying the telescope control
computer. If both arrays are read out simultaneously a potential race
condition can develop. At present, to avoid this conflict for
simultaneous observations, access to telescope information must be
disabled for one of the arrays.  This is easily done in software but
costs some header information.  However, since our observations are
sequential rather than simultaneous, this has not been necessary.

\section{Calibration \& Observing Strategy}
\label{cos}

Since the dichroic displaces the optical beam on the CCD by about
$30''$, the optical bore-sight moves relative to the IR beam depending
on whether or not the dichroic is in the beam for optical
observing. To take account of this, an offset is programmed into the
observing scripts so that a telescope motion is commanded from the CCD
data acquisition system to jog between the optical and IR beams when
observing switches from CCD-2 to LIRC-2.

VIRIS was initially deployed with an oversized transfer lens. An
unacceptably high level of scattered light, caused by a reflection
between the CCD dewar and the periphery of the transfer lens initially
made flat fields difficult to measure. In subsequent runs the
installation of an aperture stop at the lens, and application of black
material to reflective Al ring supporting the dewar window eliminated
this reflection, and permitted measurement of accurate flat fields.
The motion of the translation stage is sufficiently
reproducible
so that no observable change in flat
fields, e.g., due to dust specs on the dichroic, has been detected.

Our observational campaign has concentrated on T-Tauri stars, which
are typically red objects, requiring moderate optical exposures 20-300
s (dichroic out), and shorter (5-150 s), IR exposures. The IR
exposures consist of 5 individual, short (1-30 s), dithered frames so
that the IR bright sky does not saturate the detector and to permit
good sky subtraction.  During a typical summer night of 9 hours we can
measure some 15-20 target sources in $VRIJHK$, together with 4-6
photometric standard fields.  We primarily use stars common to the
Landolt (1983, 1992) optical standards lists and the Casali and
Hawarden (1992) infrared standards list for photometric calibration.
Table \ref{zeropts} lists the photometric zero points for VIRIS.
These data were derived from from several observations of HD 18881.
The $VR_cI_c$ data are listed for both positions of
the dichroic. 

\section{Science Results}

\begin{figure}
\plotone{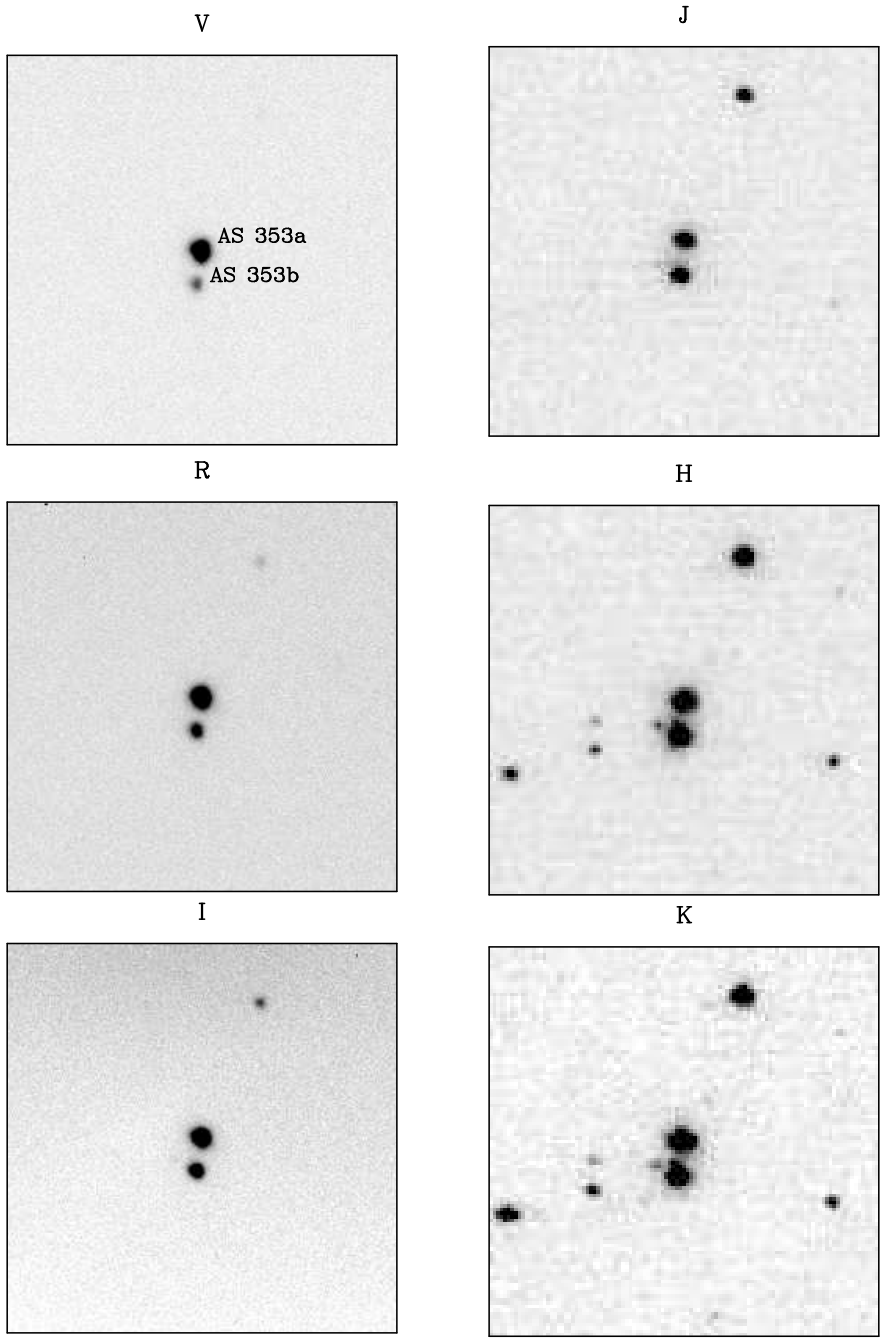} 
\vskip -0.5 in 
\figcaption{VIRIS $VRIJHK$ images of AS353 a/b, a
pre-main sequence binary system.  The field of view is $1'$ on a side,
with north up and east to the left.  The plate scale is $0.''25$ per
pixel for the optical images and $0.''57$ per pixel for the infrared
images.  Typical seeing at the Lick Observatory 1-m telescope is $\sim
1.''5$.}
\label{fig:images}
\end{figure}

\begin{figure}
\plotone{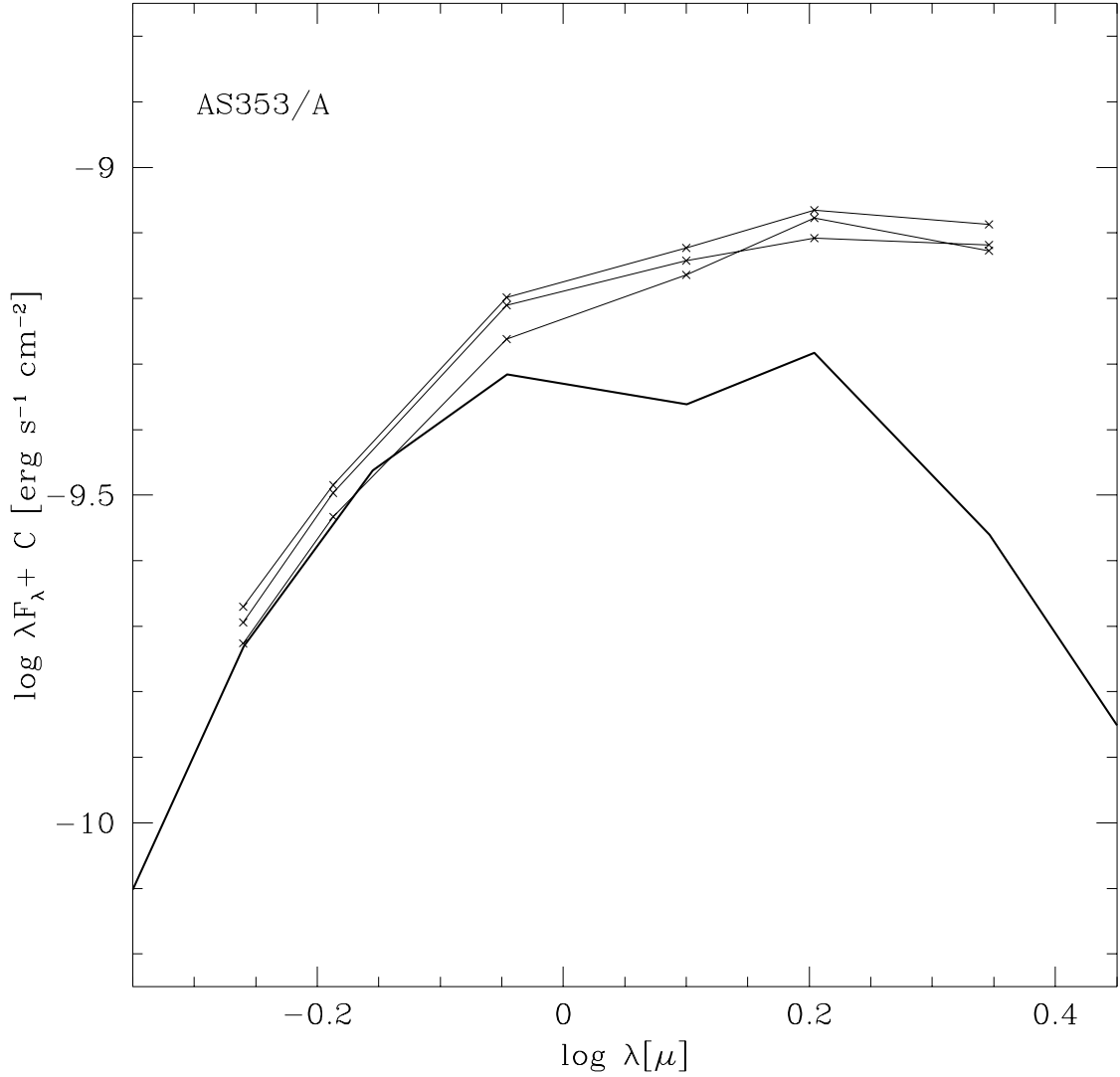}
\figcaption{Spectral energy distribution of AS353a from
0.55-2.2$\micron$.  Three consecutive nights of photometry from VIRIS
are shown ($\times$'s connected by light solid lines) along with the
spectral energy distribution of an un-reddened K5 star (heavy solid
line), for comparison.  Error bars are not shown since they would be
smaller then the plotted symbols.  The comparison star has been
arbitrarily normalized at $V$ to our faintest measurement.  Note the
flux variability at all wavelengths and the large amount of excess
emission in the near-infrared.}
\label{fig:sed}
\end{figure}

An example set of VIRIS images is shown in Figure~\ref{fig:images}
with a limited time-series spectral energy distribution shown in
Figure~\ref{fig:sed}.  The young stars we have chosen for
investigation are all $<1-2$ Myr of age, $\sim0.5-3M_\odot$ in mass,
and have demonstrated optical and near-infrared variability from data
in the literature.

Our strategy has been to observe the same list of targets every night
of every 3-5 night run with 2-4 weeks separating runs.  Thus far about
30 nights of photometry have been collected over 2 summer seasons
(1996 and 1997).  Our sampling is not good enough to search for
rotation periods, but we can correlate optical and infrared
monochromatic and color variability.  With the construction of
accurate SEDs ({\it i.e.} based on near-simultaneous optical and
infrared data), and an understanding of their variability trends, our
eventual goal is to investigate whether changes in some combination of
the geometry of the circumstellar dust and the accretion rate, can
account for the observed flux variability in these young stars.

\section{Conclusions}

We have designed and constructed an efficient system for
near-simultaneous optical and IR imaging and used it to obtain
accurate $VRIJHK$ photometry in a program to monitor the variability
of T-Tauri stars.  VIRIS demonstrates the feasibility of combining
existing optical and IR instruments using an opto-mechanical interface
to permit simultaneous and near-simultaneous operation. We have shown
that the details of the interface are very simple, and likely could be
replicated on other small telescope to permit them to tackle similar
scientific programs that require panchromatic observations.

\acknowledgments

We extend our thanks for invaluable assistance from the staff at Lick,
both in Santa Cruz and on Mt Hamilton; especially to Matt Radovan and
Rem Stone.  Support for this program was provided by Joe Miller, the
Director of Lick observatory. JRG is supported in part by a
Fellowship from the Packard Foundation. Support to LAH for this work
was provided by NASA through grant \#HF1060.01-94A from the Space
Telescope Science Institute, which is operated by the Association of
Universities for Research in Astronomy, Incorporated, under NASA
Contract NAS5-26555

\clearpage


\begin{figure}
\plotone{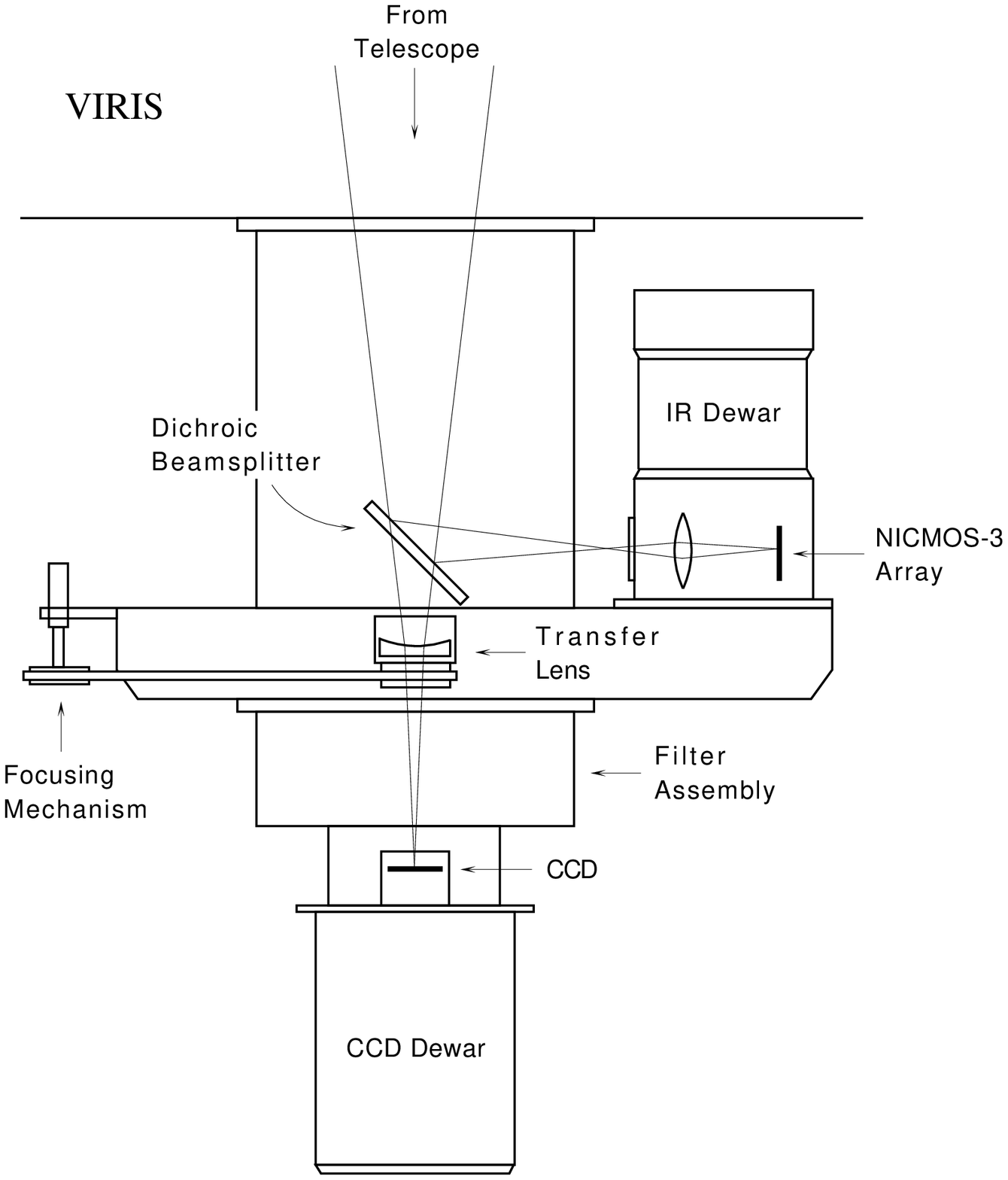}
\figcaption{Figure showing the location of CCD2 and LIRC-2 on the Lick
1-m telescope with the VIRIS interface and transfer lens}
\label{fig:setup}
\end{figure}

\clearpage

\begin{deluxetable}{lll}
\tablewidth{0pt}
\tablecaption{VIRIS Photometric Zero 
Points on the 1-m Telescope \tablenotemark{a}}
\tablehead{
\colhead{Band} & \colhead{Dichroic Out} &\colhead{Dichroic In}}
\startdata
$V$   & 22.4 & 19.9 \\
$R_c$ & 22.8 & 19.3 \\
$I_c$ & 23.5 & 19.5 \\
$J$   & ...  & 21.1 \\
$H$   & ...  & 21.1 \\
$K'$  & ...  & 20.8 \\
\enddata
\tablenotetext{a}{The zero point is defined as the
stellar magnitude which gives an integrated
signal of one photoelectron per second. 
At the time of these observations the gain of the CCD system 
($VR_cI_c$) was measured to be 6 e$^-$ per ADU and that
of the IR system ($JHK$) was  9 e$^-$ per ADU.}
\label{zeropts}
\end{deluxetable}

\end{document}